\documentclass[fleqn,usenatbib]{mnras}

\usepackage{newtxtext,newtxmath}

\usepackage[T1]{fontenc}

\DeclareRobustCommand{\VAN}[3]{#2}
\let\VANthebibliography\thebibliography
\def\thebibliography{\DeclareRobustCommand{\VAN}[3]{##3}\VANthebibliography}

\usepackage{graphicx}	
\usepackage{amsmath}	

\title[Search for transients in RELICS]{A search for transients in the Reionization Lensing Cluster Survey (RELICS): Three new supernovae}

\author[M. Golubchik et al.]{Miriam Golubchik$^{1}$\thanks{E-mail: golubmir@post.bgu.ac.il},
Adi Zitrin$^{1}$,
Justin Pierel$^{2}$,
Lukas J. Furtak$^{1}$,
Ashish K. Meena$^{1}$,
Or Graur$^{3,4}$,
\newauthor
Patrick L. Kelly$^{5}$,
Dan Coe$^{2,6,7}$,
Felipe Andrade-Santos$^{8,9}$
Maor Asif$^{1}$,
Larry D. Bradley$^{2}$,
Wenlei Chen$^{5}$,
\newauthor
Brenda L. Frye$^{10}$,
Sebastian Gomez$^{2}$,
Saurabh Jha$^{11}$,
Guillaume Mahler$^{12,13}$,
Mario Nonino$^{14}$,
\newauthor
Louis-Gregory Strolger$^{2}$,
Yuanyuan Su$^{15}$
\\
$^{1}$Physics Department,
Ben-Gurion University of the Negev, P.O. Box 653,
Be'er-Sheva 84105, Israel\\
$^{2}$Space Telescope Science Institute (STScI), 3700 San Martin Drive, Baltimore, MD 21218, USA\\
$^{3}$Institute of Cosmology \& Gravitation, University of Portsmouth, Dennis Sciama Building, Portsmouth, PO1 3FX, UK\\
$^{4}$Department of Astrophysics, American Museum of Natural History, Central Park West and 79th Street, New York, NY 10024, USA\\
$^{5}$School of Physics and Astronomy, University of Minnesota, 116 Church Street SE, Minneapolis, MN 55455, USA\\
$^{6}$Association of Universities for Research in Astronomy (AURA) for the European Space Agency (ESA), STScI, Baltimore, MD, USA\\
$^{7}$Center for Astrophysical Sciences, Department of Physics and Astronomy, The Johns Hopkins University, 3400 N Charles St. Baltimore, MD 21218, USA\\
$^{8}$Department of Liberal Arts and Sciences, Berklee College of Music, 7 Haviland Street, Boston, MA 02215, USA\\
$^{9}$Center for Astrophysics \text{\textbar} Harvard \& Smithsonian, 60 Garden Street, Cambridge, MA 02138, USA\\
$^{10}$Department of Astronomy/Steward Observatory, University of Arizona, 933 N. Cherry Avenue, Tucson, AZ 85721, USA\\
$^{11}$Department of Physics and Astronomy, Rutgers, the State University of New Jersey,
136 Frelinghuysen Road, Piscataway, NJ 08854-8019, USA\\
$^{12}$Institute for Computational Cosmology, Durham University, South Road, Durham DH1 3LE, UK\\
$^{13}$Centre for Extragalactic Astronomy, Durham University, South Road, Durham DH1 3LE, UK\\
$^{14}$INAF-Trieste Astronomical Observatory, Via Bazzoni 2, 34124, Trieste, Italy\\
$^{15}$University of Kentucky, 505 Rose Street, Lexington, KY 40506, USA}

\date{Accepted XXX. Received YYY; in original form ZZZ}

\pubyear{2023}

\begin{document}
\label{firstpage}
\pagerange{\pageref{firstpage}--\pageref{lastpage}}
\maketitle

\begin{abstract}
The Reionization Cluster Survey (RELICS) imaged 41 galaxy clusters with the \emph{Hubble Space Telescope} (HST), in order to detect lensed and high-redshift galaxies. Each cluster was imaged to about 26.5 AB mag in three optical and four near-infrared bands, taken in two distinct visits separated by varying time intervals. We make use of the multiple near-infrared epochs to search for transient sources in the cluster fields, with the primary motivation of building statistics for bright \emph{caustic crossing events} in gravitational arcs. Over the whole sample, we do not find any significant ($\gtrsim5 \sigma$) caustic crossing events, in line with expectations from semi-analytic calculations but in contrast to what may be \emph{naively} expected from previous detections of some bright events, or from deeper transient surveys that do find high rates of such events. Nevertheless, we find six prominent \emph{supernova} (SN) candidates over the 41 fields: three of them were previously reported and three are new ones reported here for the first time. Out of the six candidates, four are likely core-collapse (CC) SNe -- three in cluster galaxies, and among which only one was known before, and one slightly behind the cluster at $z\sim0.6-0.7$. The other two are likely Ia -- both of them previously known, one probably in a cluster galaxy, and one behind it at $z\simeq2$. Our study supplies empirical bounds for the rate of caustic crossing events in galaxy cluster fields to typical HST magnitudes, and lays the groundwork for a future SN rate study. 
\end{abstract}




\section{Introduction}\label{sec:intro}
Lensing of transient sources, whether theoretically or observationally, has risen in interest in recent years due to the ability to teach us about the lensed sources, but also constrain the dark matter (DM) composition of the lens (see \citealt{Oguri2019RPPh...82l6901O} for review). Among these transient sources, for example, are Fast Radio Bursts \citep[FRBs; e.g.][]{Munoz2016PhRvL.117i1301M}, Gamma Ray Bursts \citep[e.g.][]{Paynter2021NatAs...5..560P}, Supernovae \citep[SNe; e.g.][]{Kelly2015Sci,Goobar2017Sci...356..291G,Rodney2021NatAs...5.1118R}, Gravitational Waves \citep[e.g.][]{Dai2020arXiv200712709D,Broadhurst2022arXiv220205861B}, and Caustic Crossing Events of lensed stars \citep[e.g.][]{Miralda-Escude1991ApJ...379...94M,Venumadhav2017ApJCCE,Kelly2018NatAsCCE,Diego2018ApJCCE,Meena2022arXiv221113334Mz4p8}.  Among the several transients described above, the most natural transients to look for in optical and near-infrared imaging of galaxy clusters are lensed supernovae and caustic crossing events. 

Lensed SNe have been attracting much interest, in part since they are expected to be found to larger redshifts than non magnified SNe, thus potentially improving the constraints on the cosmological parameters through their contribution to the Hubble diagram. Several SNe lensed by galaxy clusters have been found to date. The first few examples were found sufficiently far from the center of the lens or at a low enough redshift behind it so they were magnified but not multiply imaged. For example, \citet{Patel2014SN,Nordin2014SN} found three lensed SNe in the 25 cluster fields of the \textit{Cluster Lensing And Supernova survey with Hubble}~\citep[CLASH;][see also \citealt{Graur2014SN}]{PostmanCLASHoverview}. In another example,  
\citet{Rodney2015ApJ...811...70R} found a Type Ia SN behind the \textit{Hubble Frontier Fields} \citep[HFF;][]{Lotz2016HFF} cluster Abell~2744, which was then confronted with magnification estimates from various lens models \citep[see also][]{Mahler2018}. The past few years have seen also the first examples of several multiply imaged SNe, long anticipated \citep{Refsdal1964MNRAS}. \emph{Refsdal}, the first multiply imaged SN, was detected by \citet{Kelly2015Sci} as an Einstein cross around a cluster galaxy in the CLASH/HFF cluster MACS~J1149.5+2223 \citep{EbelingMacs12_2007}. Another image of the SN appeared a year later in a counter-image of the SN's host galaxy \citep{Kelly2016ApJ...819L...8K}, enabling a measurement of the expansion rate of the universe from the time delays~\citep{Vega-Ferrero2018,Grillo2018ApJ...860...94G}. About a year later, another lensed SN was found, multiply imaged by a field galaxy~\citep{Goobar2017Sci...356..291G}. \citet{Rodney2021NatAs...5.1118R} have found another SN multiply imaged by a foreground massive cluster lens, MACS~J0138.0-2155 \citep{Ebeling2010MacsBrights}, with the next image predicted to appear in about two decades. More recently, \citet{Chen2022Natur.611..256C} found a multiply imaged SN around a galaxy in the HFF cluster Abell 370. At a redshift of $z \sim 3$, this SN is the farthest one known ($z\simeq3$), allowing also an early view of the SN's explosion thanks to the time delays. In addition to the physics they enable us to study, such multiply imaged transients are also important as they allow us to probe and re-calibrate our lens modeling techniques \citep[e.g.][]{Treu2016Refsdal,Zitrin2021ApJ...919...54Z}.  


The \emph{rate} of SNe in galaxy clusters has also spurred much interest \citep[e.g.][]{Gal-Yam2003SNICM,Graham2008SNrateclusters,Mannucci2008MNRAS.383.1121M,Dilday2010ApJ...715.1021D,Sharon2010SNRATE,Sand2011SNrateclusters,Maoz2012ProgenitorProblemReview,MaozGraur2017SNRate,FreundlichMaoz2021SNRate}. The process of chemical or metal enrichment by SNe is not very well constrained observationally \citep{SarkarAndSu2022}. One of the crucial ingredients to characterise it is the distribution of delay time between a burst of star formation and the explosions as SNe, especially for Type Ia -- for which the delay-time distribution spans Myr to Gyr time scales \citep{MaozGraur2017SNRate}, unlike core-collapse (CC) SNe, for example, which result from massive, short lived stars. Because most galaxies in clusters are red elliptical galaxies, most SNe we expect to see in clusters are thus Type Ia, with a higher CC to Ia fractions towards higher redshifts around when the clusters formed ($z\sim3-4$). That said, \citet{Graham2012SNIIrate}, for example, found several CC SNe even in low-redshift cluster galaxies showing that at least some low level star-formation is taking place also in these so called red-and-dead galaxies. In addition, the progenitor of Type Ia SNe is not unambiguously known. In particular, it is not clear if most Type Ia result from single or double degenerate CO white dwarfs \citep[e.g.][]{Hillebrandt2000ARA&A..38..191H}.
Since different progenitor scenarios involve different timescales that control the production rate of SN Ia events, the rate, and its dependence on the host stellar-population age, can help discriminate between these models \citep{Maoz2012ProgenitorProblemReview}. The rate of different types of SNe in cluster fields, as a function of redshift, is thus of high importance.


The detection of the first multiply imaged SN \emph{Refsdal} \citep{Kelly2015Sci} led also to the serendipitous detection of another type of transient, namely, a \emph{caustic crossing event}. In follow-up observations of the SN, a new transient was seen atop the expected position of the lensing critical curve, where the magnification gets extremely high \citep{Kelly2018NatAsCCE}. The Spectral Energy Distribution (SED) and properties of the transient matched well that of a star crossing the caustic, getting temporarily, extremely magnified \citep{Miralda-Escude1991ApJ...379...94M,Diego2018ApJCCE}. This event then opened the door to observing cosmological stars throughout the universe. Indeed, a growing number of lensed stars have been observed since \citep[e.g.][]{Chen2019CCE,Kaurov2019CCE,Meena2022arXivz1p26,Kelly2022arXivFlashlight,Diego2022}, with the highest redshift example being Earendel at $z\simeq$6.2 \citep{welch22}. JWST now offers a deeper window to observing and studying such stars through cosmic time, and has already revealed various types of stars in its first few months of operation \citep[e.g.][]{Chen2022z265star,Diego2022arXivGordo,Meena2022arXiv221113334Mz4p8,Pascale2022b,Welch2022EarendelJWST}.

In this work we report the results from a search for transients in \textit{Hubble Space Telescope} (HST) images taken for the RELICS program. In RELICS, 41 clusters were imaged in seven optical and near-infrared bands, to about 26.5 AB magnitudes per band. The four-band near-infrared (NIR) imaging was repeated in two different epochs (see section \ref{s:data}), separated by different periods of time (usually a few weeks or months), allowing us to search for transient sources. Our main motivation for the work is to estimate the rate of caustic crossing events in the survey. Although dedicated surveys with HST (e.g. the Flashlights program; \citealt{Kelly2022arXivFlashlight}) and the first clusters imaged with JWST (see works mentioned above) imply relatively high rates to $\sim29-30$ AB, the first lensed stars were discovered at a level of 26-27 AB magnitudes \citep{Kelly2018NatAsCCE,Chen2019CCE,Kaurov2019CCE}, suggesting that RELICS observations should be sufficient for detecting at least \emph{some} bright caustic crossing events. The observed rate depends, however, on various observational and physical properties such as filter choice, the background stellar mass density, the source radius-, luminosity-, and mass-functions, as well as properties of the lenses; which is why it is important to constrain.

The paper is organised as follows. We describe the data and observations in \S \ref{s:data}, and the methods in \S \ref{s:methods}. The results are presented and discussed in \S \ref{s:results}, and the work is concluded in \S \ref{s:summary}. Throughout we use a standard flat $\Lambda$CDM cosmology with $H_0=70~\,\mathrm{km}~\mathrm{s}^{-1}~\mathrm{Mpc}^{-1}$, $\Omega_{\Lambda}=0.7$, and $\Omega_\mathrm{m}=0.3$. All magnitudes quoted are in the AB system \citep{Oke1983ABandStandards} and all quoted uncertainties represent $1\sigma$ ranges unless stated otherwise. 

\begin{figure*}
 \begin{center}
 \includegraphics[width=0.98\textwidth, keepaspectratio=true]{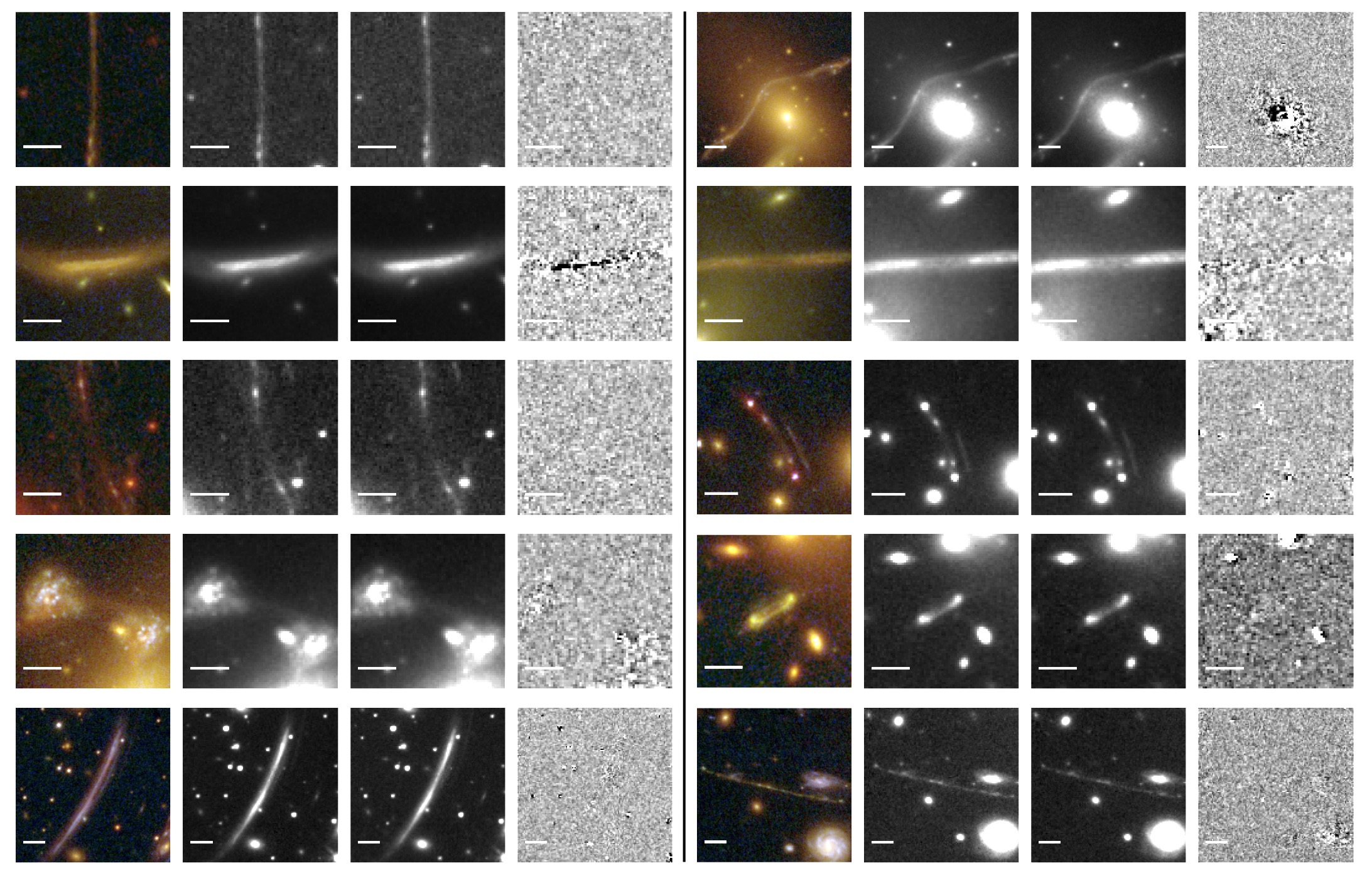}
 \end{center}
\caption{Examples of caustic crossing arcs out of 47 arcs inspected in RELICS in which one may expect to observe lensed stars. For each example arc we present, from left to right, a colour image, a combined image of all WFC3 bands for the first epoch, a similar image for the second epoch, and their difference image. The white bar in each image represents a scale of 2 arcsec. We find no clear detection of caustic crossing events in the RELICS survey. Left (from top to bottom): ACT-CL J0102–49151, A2813, ACT-CL J0102–49151, MACS J0417.5-1154, PLCK G004.5-19.5. Right (from top to bottom): AS295, A2813,  MACS J0035.4-2015, CL J0152.7–1357, MACS J0553.4-3342.}\vspace{0.1cm}
\label{fig:arcs}
\end{figure*}

\section{Observations and Data}\label{s:data}
Throughout this work we use data obtained in the framework of the RELICS program. In this program 41 massive galaxy clusters were observed with the HST (PI: D. Coe), and the \textit{Spitzer Space Telescope} (PI: M. Bradac), with the goal of detecting
gravitationally lensed arcs, bright high-redshift galaxies \citep{Salmon2020HighzRelics,strait21}, as well as various transients (e.g. see the SNe listed in \citealt{Coe2019RELICS} and the first spectacular detection of a lensed star at $z\simeq6.2$ in \citealt{welch22}). All targets were observed (from two separate epochs combined) to about 26.5\,magnitudes in 7 HST bands: F435W, F606W, F814W with the \textit{Advanced Camera for Surveys} (ACS), and F105W, F125W, F140W, and F160W, with the \textit{Wide Field Camera Three}
(WFC3). For some clusters, previous HST observations were used as well, as detailed in \citet{Coe2019RELICS}.

The RELICS data products include reduced and colour images, photometric catalogs generated with \texttt{SExtractor} \citep{BertinArnouts1996Sextractor} and photometric redshifts computed with the \texttt{Bayesian Photometric Redshifts} tool (\texttt{BPZ}; \citealp{Benitez2004,Coe2006}). These are publicly available on the RELICS website \footnote{\url{https://relics.stsci.edu/}}. We refer the reader to \citet{Coe2019RELICS} for additional details on the HST data reduction and catalog construction.

\section{Methods}\label{s:methods}
\subsection{Image Subtraction}\label{s:image_sub}
Due to the fact that the final drizzled images of each epoch in each band are not part of the RELICS data products, we assemble an image for each epoch in each of the bands using the raw `.flt' files.
For each cluster we use observations made with the WFC3 camera in four different bands (F105W, F125W, F140W, F160W). For each epoch, all exposures of the same band are aligned using \texttt{TweakReg} and combined into a final image using \texttt{AstroDrizzle} \citep[][]{Koekemoer2011}. Both functions are part of the \texttt{DrizzlePac} software package available online.
An automatic procedure is used to determine optimal values for the \texttt{`conv\_width'} and \texttt{`threshold'} parameters of the \texttt{TweakReg} function, minimizing the \emph{rms} of the offsets in x and y.
In several cases the initial offsets of the data were large and hence we used the \texttt{`search\_radius'} parameter for optimal alignment. We refer the reader to the DrizzlePac documentation\footnote{\url{https://drizzlepac.readthedocs.io/en/latest/tweakreg.html}} for further details about these functions.
The two final images are subtracted, resulting in a difference image in which transients are searched. We use once more the \texttt{AstroDrizzle} function to calculate the total errors for each epoch, and finally a signal-to-noise map is calculated. 

Note that for each cluster and for each epoch we also create a deeper ($\sim27$ AB) image by a weighted sum of the four WFC3/IR filters, and generate difference images from these ``IR-combined" images as well.

\begin{table*}
	\centering
	\caption{The six supernova candidates found in this work. \emph{Column 1:} Candidate ID, indicating also the abbreviated name of the cluster (see \S \ref{s:results} for full cluster names); \emph{Columns 2 \& 3:} RA and Dec of the supernova, in J2000.0 \emph{Column 4:}  Cluster redshift as listed in \citet{Coe2019RELICS}; \emph{Column 5:} Photo-$z$ from the RELICS catalog for the most-probable host galaxy; \emph{Columns 6 \& 7:} Dates of first and second epoch; \emph{Columns 8:} Name of supernova from \citet{Coe2019RELICS} and references therein. The three supernovae for which no names are designated are the three newly found supernovae.}
\label{multTable1}
\begin{tabular}{lccccccc}
Candidate ID & R.A. (J2000) & Dec. (J2000)&Cluster Redshift &Host $z_{\rm phot}^{a}$ &First Epoch  &Second Epoch &Name\\
\hline
RXCJ0142-SN1 &  $25.740828$ & $44.641610$  &  $0.341$ & $3.53$ 
 \ [$3.43$--$3.55$] & 2015-12-04 & 2016-01-14 & ---\\
AS295-SN1 &  $41.3927500$ & $-53.0308231$  & $0.300$ & $0.67$  \ [$0.53$--$0.72$] & 2016-08-30 & 2016-10-09 & ---\\
PLCKG171-SN1 & $48.2464374$  & $8.3786862$  & $0.270$ & $2.71$ [$2.59$--$2.76$] & 2016-09-10 & 2016-10-21 & Kukulkan\\
RXCJ0600-SN1 & $90.0510314$  &  $-20.1233171$  & $0.460$ & $0.38$ \ [$0.30$--$0.41$] & 2017-01-11 & 2017-02-15 & William\\
A1763-SN1 & $203.8130885$  & $41.0043610$  & $0.228$ & $1.75$ [$1.65$--$1.79$] & 2016-05-08 & 2016-06-17 & Nebra\\
PLCKG287-SN1 & $177.7162202$ & $-28.0932911$ & $0.390$ & $0.35$ [$0.30--0.37$] & 2017-02-21 & 2017-03-18 & ---\\
\hline
\end{tabular}
\par\smallskip
\flushleft{$^{a}$ - Note this is the host photo-$z$ as output by the RELICS pipeline. The final deduced type and redshift for each supernova, as well as an indication whether it is in or behind the cluster, is given in Table \ref{tab:lcfits} and \S \ref{s:results}.}
\end{table*}

\subsection{Transients Detection}\label{s:trans}
We search for transients in the difference images by eye. The systematic search is initially done on the F140W difference image for all RELICS clusters. Using a simple script, each difference image (typically 2'$\times$2') is split into squares with an overlap of a few arc-seconds between sub-frames. For each initial candidate identified, zoomed-in stamps in all available bands, including the IR-combined images, from both epochs and their difference images, are then generated for further inspection. Note that the search is done with a pixel scale of $\sim0.12$\arcsec. This pixel scale is comparable to the WFC3 NIR point-spread-function (PSF) and is nominally double both the RELICS pixel scale and the pixel scale we use here for photometry and measurement (see \S \ref{s:phot} below). This means that the search was made on, effectively, somewhat smoothed images, which can help in detecting some events.

We also perform another manual search on the deeper, IR-combined images using \texttt{SAOImageDS9}, especially around gravitationally lensed arcs, in a more focused attempt to detect caustic crossing events. In practice we go over 47 arcs over all clusters (see Fig. \ref{fig:arcs} for examples). The mean photometric redshift of the arcs is $z\sim1.8$, with a standard deviation of $\sim1.2$.

The above searches are made on both the difference image and on its negative, to minimise biases related to white-versus-black detections. For each cluster we then perform a more detailed inspection of all candidates that survived the previous steps. 
We accept a candidate as a reliable transient if it appears in the difference images of all four NIR bands and in the relevant epoch in all available bands, and corresponds in practice to a signal-to-noise ratio higher than about $\sim5$. This level was estimated by planting some point sources with different signal-to-noise ratios in some images, convolved with the PSF, and repeating the detection process. We also check whether the transient appears in RELICS colour images. In the scrutinizing process, the shape of the transient is also considered, as well as its location in the field, e.g., some events in the very edges were discarded as likely artefacts.

\subsection{Photometry measurements}\label{s:phot}
We register our WFC3/IR images for the two epochs, and the RELICS optical images to the \textit{Gaia} DR3 \citep{GaiaDR32022} astrometry using \texttt{Scamp} \citep{Bertin2006SCAMP} and re-sample the images with \texttt{Swarp} \citep{BertinSwarp} onto the same 0.06\arcsec/pix grid as is used in RELICS. The photometry is measured on these images with the \texttt{photutils} package \citep[\texttt{v1.5.0};][]{photutils22} in a circular aperture, with a radius of $\sim3$ pixels of $0.06\arcsec$, and corrected for local background flux measured in a circular annulus around the source. Exact apertures and background radii were slightly refined manually for each candidate, to minimise contamination by the host galaxy. We also run the same photometry on the difference images themselves. This constitutes an important consistency check as the difference in flux between the photometry of the first and second epochs should match the fluxes measured in the difference image. All photometry is summarised in Table \ref{multTable2}. The images, including the difference images, as well as the photometry with relevant Julian date for each band, are made available online\footnote{\url{https://www.dropbox.com/scl/fo/5yfm78d1kwizylh01gone/h?dl=0&rlkey=k00fqpuj4yynamatfq182l17a}}.

\subsection{Supernovae light curve fitting}\label{s:lcfit}
\begin{figure*}
    \centering
    \includegraphics[width=\linewidth,trim={1cm 1cm .5cm 0cm}, clip]{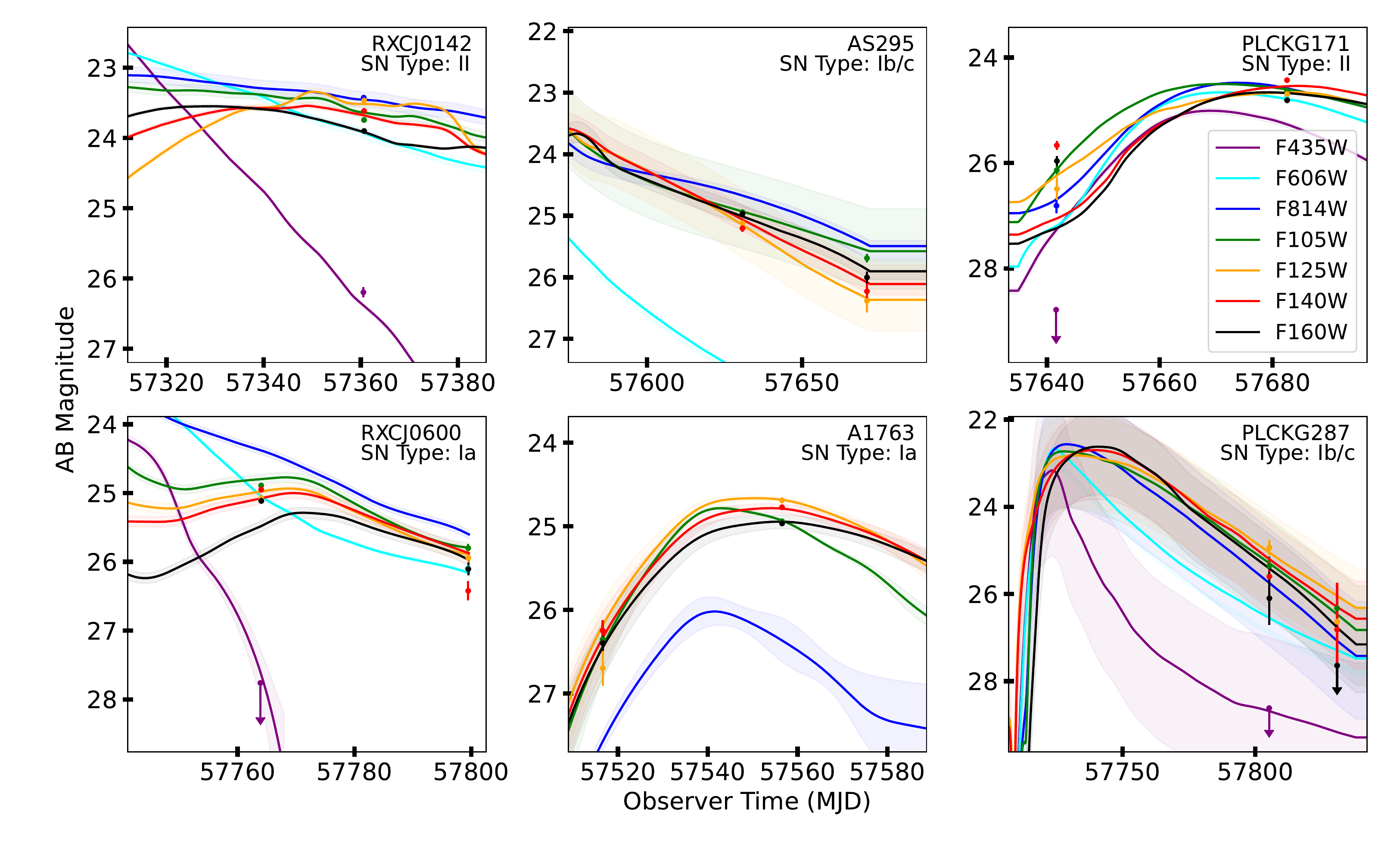}
    \caption{The best fit model for each SN light curve. Each colour corresponds to a different \textit{HST} filter, with measurements shown as points with errors (or upper-limit arrows) and maximum likelihood model realizations shown as solid lines. Shaded regions of the same colour correspond to $1\sigma$ uncertainties from the fitting procedure. The SN type shown is the same as in Table \ref{tab:lcfits}, and corresponds to each plotted model.}
    \label{fig:lcfits}
\end{figure*}

We attempt to classify each of the SN candidates using the STARDUST2 Bayesian light curve classification
tool \citep{Rodney2014Candels}, which is built on the underlying \texttt{SNCosmo}
framework and designed for classifying SNe using \textit{HST}. STARDUST2 uses
the SALT3-NIR model to represent Type Ia SNe
\citep{Pierel2022SALT3} and a collection of 42 spectrophotometric time
series templates to represent CC SNe (27 Type II and
15 Type Ib/c). These CCSN templates comprise all
of the templates developed for the SN analysis software \texttt{SNANA} \citep{Kessler2009SNANA}, derived from the SN samples of
the Sloan Digital Sky Survey \citep{Frieman2008SDSS,Sako2008SDSS,DAndrea2010SDSS}, \textit{Supernova
Legacy Survey} \citep{Astier2006SNLS}, and \textit{Carnegie Supernova Project}
\citep{Hamuy2006CSP,Stritzinger2009CSP,Morrell2012CSP}, and extended to the NIR by \citet{Pierel2018Templates}. With STARDUST2 we use a nested sampling
algorithm to measure likelihoods over the SN simulation parameter space, including priors on dust parameters described in \citet{Rodney2014Candels}. Nested sampling is a Monte Carlo
method that traverses the likelihood space in a manner
that samples the Bayesian likelihood \citep{Skilling2004NEST}. The results of the fitting and classification procedure are summarised by Table \ref{tab:lcfits} and Fig. \ref{fig:lcfits}. We find two likely SNe\,Ia, one in a cluster member and one the previously discovered SN \textit{Nebra} (see Table \ref{multTable1}), two CC\,SNe with $75\%$ and $84\%$ probabilities of being SNe\,Ib/c, and two likely SNe\,II. Of the CC\,SNe, all but one appear to be cluster members. These are further detailed in the results section \S \ref{s:results}.

\begin{table*}
	\centering
	\caption{Summary of the SN light-curve fits. \emph{Column 1:} ID; \emph{Column 2:} Redshift of the cluster; \emph{Column 3:} Redshift range allowed in the light-curve fitting. In case the fit was performed using the cluster's redshift a single value is shown. \emph{Column 4:} Best-fit redshift from the light-curve fit, in case a range was allowed; \emph{Column 5:} SN classification; \emph{Column 6:} Probability of each SN type; \emph{Column 7:} Time of peak (MJD); \emph{Column 8:} Reduced chi-square; \emph{Column 9:} Absolute B-band magnitude. For further details on the light-curve fitting procedure see section \ref{s:lcfit}.}
\label{tab:lcfits}
\begin{tabular}{lcccccccc}
&&&&&Probability   \\
Candidate ID & Cluster $z$  & $z$ bounds & Fitted $z$   &Classification   & Ia--II--Ib/c & Peak & Reduced chi-square & Absolute mag (B-band)\\
\hline
RXCJ0142-SN1 &  $0.341$ & $0.341$ &--& II & $0.0$--$1.0$--$0.0$ & $57305$ & $11.2$ & $-17.2$ \\
AS295-SN1 &  $0.300$ &[$0.53$--$0.72$] &$0.64\pm0.02$& Ib/c & $0.25$--$0.0$--$0.75$ & 57542 & $1.2$ & $-18.4$ \\
PLCKG171-SN1 & $0.270$ & $0.270$ &--& II & $0.0$--$1.0$--$0.0$ & 57671 & $13.2$ & $-16.0$\\
RXCJ0600-SN1 & $0.460$ & $0.460$ &--& Ia & $1.0$--$0.0$--$0.0$ & 57727 & $9.6$ & $-18.9$\\
A1763-SN1 & $0.228$ & [$1.65$--$2.0$] &$1.96\pm0.04$& Ia & $0.79$--$0.21$--$0.0$ & 57553 & $1.0$ & $-20.1$\\
PLCKG287-SN1 & $0.390$ & $0.390$ &--& Ib/c & $0.0$--$0.23$--$0.77$ & 57728 & $0.6$ & $-18.3$\\
\hline
\end{tabular}
\end{table*}


\section{Results and Discussion}\label{s:results}
In our transient search we find no strong caustic crossing event candidates. This may be surprising at first sight, given that the first couple of lensed stars were detected to roughly similar magnitudes as those reached by RELICS. However, as we show below in section \ref{s:cce}, this is in broad agreement with a rate expectation based on an order-of-magnitude calculation using some simple assumptions. We do find, however, six other prominent transients, likely SNe, detailed in section \ref{s:scands}.

\subsection{Caustic Crossing Events}\label{s:cce}
We now briefly estimate the expected rate of events one should detect to the depth reached by RELICS, over 41 cluster fields. 

To do this, we start by noting that caustic crossing events are biased towards bright and luminous stars. For an O/B-type star at redshift $z=2$, with effective temperatures of $T_{\rm eff}=12,000-45,000$~K, a magnification in the range $\sim20,000-50,000$ is needed for it to be visible at an apparent magnitude of~$26.5-27.0$~AB in the HST filters. In the corrugated network forming around the macro-critical curve, thanks to point masses in the lens such as stars etc., the typical peak magnification for a stellar source of radius $R$ is expected to be \citep[e.g.,][]{Venumadhav2017ApJCCE,Oguri2018PhRvDCCE}~$[10^4,10^5]\times(R/10{\rm R_\odot})^{-1/2}$, with the exact value depending on the macro-convergence, micro-lens density, micro-lens mass function etc., with an average frequency of peaks of about $\sim1$ per year, per source star. For the above we assume a typical stellar mass density of 10 M$_{\odot}$/pc$^{2}$. For a typical lensed arc we adopt a typical apparent brightness of $\sim26~{\rm AB/arcsec^2}$. This is equivalent to $\sim10^9~{\rm L_\odot/arcsec^2}$ for a source at $z\sim2$, which roughly translates into $\sim 1,000$ massive stars per arcsec$^2$. For this order-of-magnitude calculation we assume a Salpeter IMF ranging between 0.1 and a 100 M$_{\odot}$. Given most lensed arcs are typically blue star-forming galaxies, we neglect here the time evolution of the stellar population. It should however be acknowledged that the age, as well as other factors such as metallicity, may be important for assessing the true number of expected stars, and our calculation remains crude in that sense. In particular, a more evolved stellar population will include less massive stars so our estimation is in that aspect, an upper limit. The typical relevant area of the corrugated network in the image plane is~0.1~arcsec$^2$, equivalent to 0.0005~arcsec$^2$ in the source plane, assuming a magnification of 200 within the corrugated network. Assuming a velocity of 500 km s$^{-1}$, this implies that we expect roughly one crossing every two years in an arc. 

Assuming that these events follow a Poissonian distribution, with an event lasting for~$\sim3$~days, we can estimate the probability of detecting at least one event in a given arc, in one visit, to be $\sim0.4\%$. The probability of detecting at least one event over all the RELICS clusters, assuming one caustic-crossing arc per cluster, is~$\sim16\%$. Note that this is in principle, an upper limit: stars in the corrugated network that are sufficiently close to the main caustic, such that they are constantly sufficiently highly magnified, will only show minor fluctuations and will not be detected as transient sources in our difference images. 


Outside the corrugated network, the magnification needed for O/B-type stars is typically too high to be reached in an individual micro-caustic crossing. Still, such a magnification can be reached if several microlenses sit near each other so that we have overlapping micro-caustics. Mircolensing simulations \citep{Meena2022MNRAS.514.2545M}, show that the frequency of bright enough events in this regime is 1-2 orders of magnitude lower than on the corrugated network, hence contributing only little to the chances of seeing an event in RELICS. 

Note however that some stars, such as various super- and hyper-giants, may be intrinsically brighter than what we assume above, so they could be observed even with an almost order-of-magnitude lower magnification than considered here. One such example is Icarus, for which a peak magnification of $\sim3000$ was sufficient for its detection. This means that in practice, there may be somewhat more events expected than estimated above, although due to their low numbers (or the small area in which such magnifications can be reached), we do not expect a significant contribution from such stars.

\subsection{SN candidates}\label{s:scands}
Our study yields 6 strong SN candidates summarised in Table \ref{multTable1}. Three of the candidates were previously reported in \citet{Coe2019RELICS}. We present the SNe candidates in Fig. \ref{fig:curve_all}. Four of the transients are detected to be fading and two are getting brighter between the first and second epoch. The SN light-curve fits are summarised in Table \ref{tab:lcfits} and shown in Fig. \ref{fig:lcfits}. While in principle SNe in galaxy clusters can appear in the diffuse intracluster light, all six SNe we find here seem to lie in potential host galaxies. We in addition use the \texttt{BayEsian Analysis of GaLaxy sEds} tool \citep[\texttt{BEAGLE};][]{Chevallard2016} to infer the properties of the host galaxies based on their broad-band photometry presented in Table~\ref{multTable2}. Note that we assume the same redshifts for both the light-curve fitting of the SNe and the SED-fitting of their host galaxies.

\subsubsection{RXCJ0142.9+4438}\label{s:cand1}
We detect a transient in the cluster RXC~J0142.9+4438 ($z = 0.341$; Fig. \ref{fig:curve_all}). The transient is detected to be fading between the first and the second epoch. The apparent host has a photometric redshift of $z_{phot} \approx 3.53$ in the RELICS catalog. If this redshift is correct, this would make the SN the farthest one detected by \emph{HST}. Indeed, the host galaxy does seem to be bluish-greenish in the RELICS colour image  (Fig. \ref{fig:curve_all}) suggesting it is not a typical cluster member, but possibly behind the cluster. Our lens model implies that the galaxy would not be multiply imaged at that redshift, and indeed we do not identify any possible multiple images. The host galaxy does not seem to be stretched tangentially, as would be expected by the lensing shear in this case, thus disfavoring the high-redshift solution. For second redshift estimate, we also run an \texttt{EAZY} \citep{Brammer2008EAZY} photometric-redshift fit, using the RELICS photometry for the host. We obtain a lower redshift solution, similar to the cluster redshift, namely a best-fit redshift and 68\% confidence interval of 0.34 [0.26--0.43], which would suggest the host is in the cluster. The SN light curve fit is thus run with the cluster's redshift as input (see also Table \ref{tab:lcfits}). The photometry fits well a CC SN of Type II, with more than 90\% probability. While future spectroscopy of the host would be useful for securing the host's redshift, we conclude that it is likely a CC SN of Type II at the redshift of the cluster. The SN was not known before and reported here for the first time, to the best of our knowledge. As for the host galaxy, we find it is relatively massive with $\log(M_{\star}/\mathrm{M}_{\odot})=9.50\pm0.02$, relatively young with $t_{\mathrm{age}}\simeq270$\,Myr, and has a moderate star-formation rate (SFR) of $\log(\psi/\mathrm{M}_{\odot}\,\mathrm{yr}^{-1})=0.3\pm0.2$.

\subsubsection{Abell S295}
In the cluster Abell S295 ($z = 0.300$) we detect a transient fading between the first and the second epoch. The transient appears to be inside a host galaxy (see Fig. \ref{fig:curve_all}) with a photometric redshift of $z_{\rm phot} \approx 0.67$. A photometric-redshift \texttt{EAZY} fit yields a similar redshift of 0.67, with a narrow 68\% confidence interval of [0.61--0.71]. From the light curve fit, in which the SN's redshift is free to vary between $\sim0.5$ and $\sim0.7$, we obtain it is likely (with 75\% probability) a CC of Type Ib/c, at an approximated redshift of $0.64\pm0.02$. The SN was not known before and reported here for the first time, to our knowledge. Our SED fit to the host galaxy suggests a massive ($\log(M_{\star}/\mathrm{M}_{\odot})=10.4\pm0.1$), relatively young ($t_{\mathrm{age}}\simeq160$\,Myr), dusty ($A_V=3.3_{-0.3}^{+0.4}$) and heavily star-forming spiral with an SFR of $\log(\psi/\mathrm{M}_{\odot}\,\mathrm{yr}^{-1})=1.5\pm0.3$.

\subsubsection{PLCK G171.9-40.7}
The transient detected in PLCK G171.9-40.7 ($z = 0.270$) gets brighter between the first and the second epoch and appears to lie inside a very faint host galaxy, with $z_{\rm phot} \approx 2.71$ (see Fig. \ref{fig:curve_all}). Based on other multiple images known in this cluster \citep{Cerny2018}, we would expect the host to be multiply imaged, were it at $z\sim2$ or above. However, we do not detect any counter image in the expected position based on the lensing symmetry, which suggests the galaxy is probably at a lower redshift. Preliminary investigation of this candidate in \citep{Coe2019RELICS} classified the host as a cluster member galaxy.
We run an \texttt{EAZY} photometric-redshift fit to the host photometry in the first epoch, where the SN's contribution is negligible, and obtain a relatively wide range of possible redshifts, with a single-template best-fit and 68\% confidence interval of 0.737 [0.05--2.11], leaving the redshift ambiguous. For this redshift, our lens model suggests a magnification of $\mu\sim2.3$. Given the wide photometric redshift range, for simplicity we run the light-curve fit assuming the cluster's redshift. The fit suggests that it is a CC SN of Type II, with over 90\% probability. For the host galaxy, we find a low stellar mass of $\log(M_{\star}/\mathrm{M}_{\odot})=7.58_{-0.07}^{+0.08}$, a very low SFR of $\log(\psi/\mathrm{M}_{\odot}\,\mathrm{yr}^{-1})=-1.61\pm0.3$ and a very young stellar age of $t_{\mathrm{age}}\simeq13$\,Myr. Note however that given the faintness of the object and the relatively large uncertainties on the photometry, \texttt{BEAGLE} does not find a very good fit at the cluster redshift assumed for the light-curve fit of the SN. In addition, since it is fit at a very low redshift, this galaxy does not have any rest-frame UV photometry to properly constrain its current population of massive stars which produce CC SNe.

\subsubsection{RXC J0600.1-2007}
The transient detected in the cluster RXC~J0600.1-2007 ($z = 0.460$) seems to be fading between the first and the second epoch and appears to have exploded in the outskirts of a host galaxy (see Fig. \ref{fig:curve_all}) with a BPZ photometric redshift and 95\% confidence interval of 0.383 [0.30--0.41], not too far from the cluster's redshift. This candidate has been previously reported in \citet{Coe2019RELICS}, where it was mentioned that the host was probably a cluster member. We perform the light-curve fit using the cluster redshift as input and obtain that this is likely a Type Ia SN, with more than 90\% probability. The host galaxy is a massive ($\log(M_{\star}/\mathrm{M}_{\odot})=10.96\pm0.02$), quiescent ($\log(\psi/\mathrm{M}_{\odot}\,\mathrm{yr}^{-1})=0.1_{-1.3}^{+0.5}$) cluster galaxy of $t_{\mathrm{age}}\simeq960$\,Myr.

\subsubsection{Abell 1763}
We detect a transient in the cluster Abell 1763 ($z = 0.228$), which seems to get brighter between the first and the second epoch, and appears to lie inside a very faint host galaxy ($z_{\rm phot} \approx 1.75$) see Fig. \ref{fig:curve_all}. This SN had been followed up with further observations, and was reported in some more detail in \citet{Rodney2016ATel.9224....1R,Coe2019RELICS} where it was classified as a lensed Type Ia SN at $z\sim2$. Initial magnification estimate yielded $mu\sim2$ (private communication). In this work, similar to all other cluster fields, we only use the first two epochs for our detection and analysis. Nevertheless, the light-curve fit, allowing a redshift range of [1.65--2] and based only on those two epochs, suggests a Type Ia SN at a redshift of $z\simeq2$, with 80\% probability and in agreement with the previous analysis. We find the host galaxy has a relatively high stellar mass of $\log(M_{\star}/\mathrm{M}_{\odot})=9.0\pm0.2$ and a low SFR $\log(\psi/\mathrm{M}_{\odot}\,\mathrm{yr}^{-1})=-0.7_{-0.9}^{+1.0}$ and stellar age of $t_{\mathrm{age}}\simeq120$\,Myr.

\subsubsection{PLCK G287.0+32.9}
We detect a transient in the cluster PLCK G287.0+32.9 ($z = 0.390$). The transient appears to be fading between the first and the second epoch and apparently lies in the halo of a cluster member host galaxy with $z_{phot} = 0.35~[0.30-0.37]$ (see Fig. \ref{fig:curve_all}). This candidate has not been previously reported, to our knowledge. We adopt the redshift of the cluster for the light-curve fit and obtain that this is most likely a Type Ib/c CC SN (with over 75\% probability). In this case, the host galaxy is a very massive ($\log(M_{\star}/\mathrm{M}_{\odot})=11.76\pm0.01$), quiescent ($\log(\psi/\mathrm{M}_{\odot}\,\mathrm{yr}^{-1})=-1.1_{-0.6}^{+0.7}$) and old ($t_{\mathrm{age}}\simeq10$\,Gyr) cluster member galaxy.

\section{Conclusion}\label{s:summary}
In this work we present results from a search for transients in the 41 RELICS cluster fields. Motivated by the discovery of various caustic crossing events of lensed stars in imaging depths broadly similar to those obtained in RELICS \citep[e.g.,][]{Kelly2018NatAsCCE,Chen2019CCE}, and the rapidly increasing numbers of events detected in deeper observations \citep[e.g.;][]{Kelly2022arXivFlashlight,Meena2022arXiv221113334Mz4p8,Pascale2022b}, our main goal was to characterise the caustic-crossing event rate in lensed arcs in the RELICS survey.

We utilise the fact that the NIR imaging took place in two distinct epochs to create difference images for each cluster and search for transients. We do not detect any prominent caustic-crossing event. We calculate the expected rate of events given the observational depth and find that the probability to detect at least one caustic crossing event in a cluster is $\sim0.04\%$ per visit. Assuming one lensed arc per cluster we finally conclude a probability of $\sim16\%$ to detect at least one caustic crossing event in RELICS. Our study thus supplies an empirical limit on the rate of bright caustic crossing events, to typical HST magnitudes, suggesting that indeed, a depth of $\sim26.5$ AB is insufficient for detecting substantial numbers of lensed stars and that deeper observations are needed (as indeed successfully demonstrated in the references above).

Note that some lensed stars may also appear as quasi-persistent sources, in case they are sufficiently close to the caustic, in the so-called corrugated caustic network. One such famous example is Earendel, the farthest known lensed star detected at redshift $z\simeq6.2$ whose observed brightness has remained approximately constant over several years \citep{welch22}, with only mild fluctuations. As another example, some types of stars such as Luminous Blue Variables can be bright enough for long periods of time at redshifts of $z\gtrsim1$, even if farther away from the caustic. One such prominent example is \textit{Godzilla} in the Sunrise Arc \citep{Diego2022}. Since these are not expected to appear as transient sources we also search for possible point-sources near to where the critical curves pass in lensed arcs. We find no additional, prominent point sources in caustic-crossing arcs in RELICS, although we note also that this perhaps merits a further, designated examination.

While we do not find any lensed stars, our search yields six SN candidates. Three were previously known and reported in \citet{Coe2019RELICS}, and three are, to the best of our knowledge, new ones presented here for the first time. Note also that \citet{Coe2019RELICS} found a few other candidates that are however mostly smaller or fainter-looking than our candidates, and not retained in our search (such detection differences can be attributed to different image-subtraction and source-identification procedures). The SN candidates we find here are classified using
the Bayesian light-curve fitting code STARDUST2, and we obtain that two of the six SNe are Type Ia candidates, and four are core collapse: two Type II candidates, and two candidates that are most probably Type Ib/c. Four of the SNe go off in the cluster, likely in cluster galaxies, where two seem to be lensed and lie behind the cluster, the farthest of them at $z\simeq2$. In terms of type, three of the four SNe found inside the clusters are CC, and one is Ia. This ratio is perhaps a bit surprising given that the delay times from stellar formation to explosion is much longer for Ia than for CC, and that cluster galaxies are relatively early type, so-called red-and-dead galaxies with very little new star-formation. This may be partially explained by the fact that some of these SN host galaxies do not seem to be typical cluster members (i.e., red passive ellipticals). Additionally, the SZ-selected RELICS sample may be biased towards relatively massive, younger clusters in which galaxies possibly have some more star-formation compared to the average cluster galaxy at similar redshifts. We run \texttt{BEAGLE} to extract the physical parameters of each host, and obtain that for three of the CC SN hosts the SFR is between a few and a few dozen M$_{\odot}$/yr, although for the fourth it is very low ($\sim0.03$ M$_{\odot}$/yr).  For the two Ia SN hosts, the SFR is between $\sim0.1$ and $\sim3$ M$_{\odot}$/yr, i.e., on average lower than for the CC hosts, as may be expected. Given the low number of SNe in our sample, we however do not attempt to draw general conclusions regarding this ratio and the link to the host's properties.

The SN rate in galaxy clusters, particularly as a function of redshift and type, is of high importance for a variety of studies from characterising the metal enrichment history in the cluster, through estimating the quenching and ICL distribution timescales {\citep{MaozGraur2017SNRate}, to shedding light on the SN progenitor problem \citep{Maoz2012ProgenitorProblemReview}. The detection of several SNe in this work thus calls for a rate calculation. In a follow-up work we aim to estimate the completeness of SNe detection in our method, extend the search to the parallel fields accompanying each RELICS cluster field, and calculate the resulting SN rate.

\vspace{-0.4cm}
\section*{acknowledgements}
We wish to thank the RELICS collaboration for data products that enabled this work. The BGU group acknowledges support by Grant No. 2020750 from the United States-Israel Binational Science Foundation (BSF) and Grant No. 2109066 from the United States National Science Foundation (NSF), and by the Ministry of Science \& Technology, Israel. P.L.K. acknowledges support through NSF grant AST-1908823.

This work is based on observations made with the NASA/ESA Hubble Space Telescope obtained from the Space Telescope Science Institute, which is operated by the Association of Universities for Research in Astronomy, Inc., under NASA contract NAS 5–26555. These observations are associated with program ID 15959. Support for program 15959 was provided by NASA through a grant from the Space Telescope Science Institute, which is operated by the Association of Universities for Research in Astronomy, Inc., under NASA contract NAS 5-26555. This work is also based on observations made with ESO Telescopes at the La Silla Paranal Observatory obtained from the ESO Science Archive Facility.

This research made use of \texttt{Astropy},\footnote{\url{http://www.astropy.org}} a community-developed core Python package for Astronomy \citep{astropy13,astropy18} as well as the packages \texttt{NumPy} \citep{vanderwalt11}, \texttt{SciPy} \citep{virtanen20}, \texttt{matplotlib} 
\citep{hunter07}, \texttt{specutils} \citep[][]{specutils21}, \texttt{spectral-cube} \citep{spectral-cube14} and some of the astronomy \texttt{MATLAB} packages \citep{maat14}. 

\clearpage

\begin{table*}
	\centering
	\caption{Aperture photometry for the six SN candidates found in this work and their host galaxies. For each SN we list the photometry in both the first epoch and the second epoch, as well as the photometry on the difference images in the near-IR for consistency. For the host galaxy we use the default (isophotal) photometry from the RELICS catalogs. In cases where the transient is not deteceted we note a lower bound corresponding to $1\sigma$ point-source AB magnitude limit.}
\label{multTable2}
\begin{tabular}{lccccccc}
\hline
\hline
~~~~~~~~AB magnitude in: & F435W  & F606W & F814W   & F105W   & F125W & F140W & F160W\\
\hline
\textbf{RXCJ0142-SN1}\\
Difference Image & --             & --               & --               &  $24.42\pm0.02$          &  $24.20\pm0.03$          & $24.29\pm0.03$          & $24.91\pm0.03$           \\
Epoch 1          & $26.19\pm0.07$ & --               & $23.42\pm0.01$   &  $23.74\pm0.02$   &  $23.46\pm0.02$   & $23.61\pm0.02$   & $23.90\pm0.03$    \\
Epoch 2          & --             & $24.76\pm0.02$   & --               &  $24.39\pm0.04$   &  $24.25\pm0.04$   & $24.48\pm0.06$   & $24.40\pm0.04$    \\
\hline
\textbf{RXCJ0142-SN1-Host}  \\ 
RELICS Data      & $23.34\pm0.03$ & $22.23\pm0.01$   & $21.63\pm0.01$   &  $21.63\pm0.01$   &  $21.63\pm0.01$   & $21.56\pm0.02$   & $21.50\pm0.01$    \\
\hline
\hline
\textbf{AS295-SN1}\\
2015 Jan. 23     & $>28.95$       & $>29.35$         & $>28.17$         &  --               &  --               & --               & --                \\
Difference Image & --             & --               & --               &  $25.72\pm0.06$          &  $25.63\pm0.09$          & $25.74\pm0.08$          & $25.30\pm0.04$           \\
Epoch 1          & --             & --               & --               &  $24.94\pm0.04$   &  $25.12\pm0.06$   & $25.20\pm0.06$   & $24.97\pm0.04$    \\
Epoch 2          & --             & --               & --               &  $25.69\pm0.07$   &  $26.38\pm0.19$   & $26.23\pm0.14$   & $26.00\pm0.09$    \\
\hline
\textbf{AS295-SN1-Host}  \\
RELICS Data      & $22.33\pm0.02$ & $21.64\pm0.01$   & $20.85\pm0.01$   &  $20.59\pm0.01$   &  $20.46\pm0.01$   & $20.41\pm0.01$   & $20.36\pm0.01$    \\
\hline
\hline
\textbf{PLCKG171-SN1}  \\
Difference Image & --             & --               & --               &  $24.94\pm0.04$          &  $24.81\pm0.06$          & $24.84\pm0.05$          & $25.23\pm0.06$          \\
Epoch 1          & $28.78\pm0.94$ & --               & $26.81\pm0.15$   &  $26.13\pm0.10$   &  $26.49\pm0.21$   & $25.66\pm0.09$   & $25.96\pm0.09$    \\
Epoch 2          & --             & $24.77\pm0.02$   & --               &  $24.63\pm0.03$   &  $24.69\pm0.05$   & $24.42\pm0.04$   & $24.81\pm0.04$    \\
\hline
\textbf{PLCKG171-SN1-Host}  \\ 
RELICS Data      & $23.40\pm0.13$ & $22.65\pm0.03$   & $23.05\pm0.01$   &  $22.94\pm0.04$   &  $22.74\pm0.05$   & $22.70\pm0.04$   & $22.74\pm0.03$    \\
\hline
\hline
\textbf{RXCJ0600-SN1}  \\
Difference Image & --             & --               & --               &  $25.49\pm0.05$          &  $25.67\pm0.10$          & $25.26\pm0.06$          & $25.67\pm0.06$           \\
Epoch 1          & $27.76\pm0.26$ & --               & $26.93\pm0.12$   &  $24.89\pm0.03$   &  $25.08\pm0.06$   & $24.95\pm0.04$   & $25.11\pm0.04$    \\
Epoch 2          & --             & $26.09\pm0.04$   & --               &  $25.80\pm0.06$   &  $25.94\pm0.11$   & $26.42\pm0.14$   & $26.10\pm0.09$    \\
\hline
\textbf{RXCJ0600-SN1-Host}  \\ 
RELICS Data      & $22.34\pm0.03$ & $20.514\pm0.004$ & $19.602\pm0.002$ &  $19.204\pm0.003$ &  $18.990\pm0.004$ & $18.850\pm0.003$ & $18.725\pm0.002$  \\
\hline
\hline
\textbf{A1763-SN1}  \\
Difference Image & --             & --               & --               &  $25.26\pm0.04$          &  $24.92\pm0.06$          & $25.13\pm0.06$          & $25.33\pm0.04$           \\
Epoch 1          & $26.78\pm0.15$ & --               & $27.38\pm0.19$   &  $26.35\pm0.09$   &  $26.70\pm0.21$   & $26.25\pm0.12$   & $26.40\pm0.09$    \\
Epoch 2          & --             & $26.97\pm0.09$   & --               &  $24.94\pm0.03$   &  $24.69\pm0.04$   & $24.77\pm0.03$   & $24.97\pm0.03$    \\
\hline
\textbf{A1763-SN1-Host}  \\
RELICS Data      & $24.72\pm0.06$ & $24.75\pm0.04$   & $24.75\pm0.06$   &  $24.22\pm0.05$   &  $23.55\pm0.02$   & $23.76\pm0.06$   & $23.70\pm0.02$    \\
\hline
\hline
\textbf{PLCKG287-SN1}  \\
2016 Aug. 3      & --             & $>29.35$         & $>27.99$         &  --               &  --               & --               & --                \\
Difference Image & --             & --               & --               &  $25.96\pm0.11$         &  $25.22\pm0.07$          & $25.98\pm0.12$          & $26.50\pm0.13$         \\
Epoch 1          & $28.62\pm0.66$ & --               & --               &  $25.36\pm0.22$   &  $24.93\pm0.18$   & $25.60\pm0.34$   & $26.10\pm0.61$    \\
Epoch 2          & --             & --               & --               &  $26.33\pm0.45$   &  $26.63\pm0.84$   & $26.82\pm1.07$   & $27.64\pm2.56$    \\
\hline
\textbf{PLCKG287-SN1-Host}  \\
RELICS Data      & $21.06\pm0.02$ & $20.67\pm0.01$   & $19.435\pm0.003$ &  $18.575\pm0.002$ &  $18.194\pm0.002$ & $18.045\pm0.001$ & $17.931\pm0.003$  \\
\hline
\hline
\end{tabular}
\end{table*}

\clearpage

\begin{figure*}
 \begin{center}
\includegraphics[height=0.40\textwidth,width=0.49\textwidth, keepaspectratio=true]{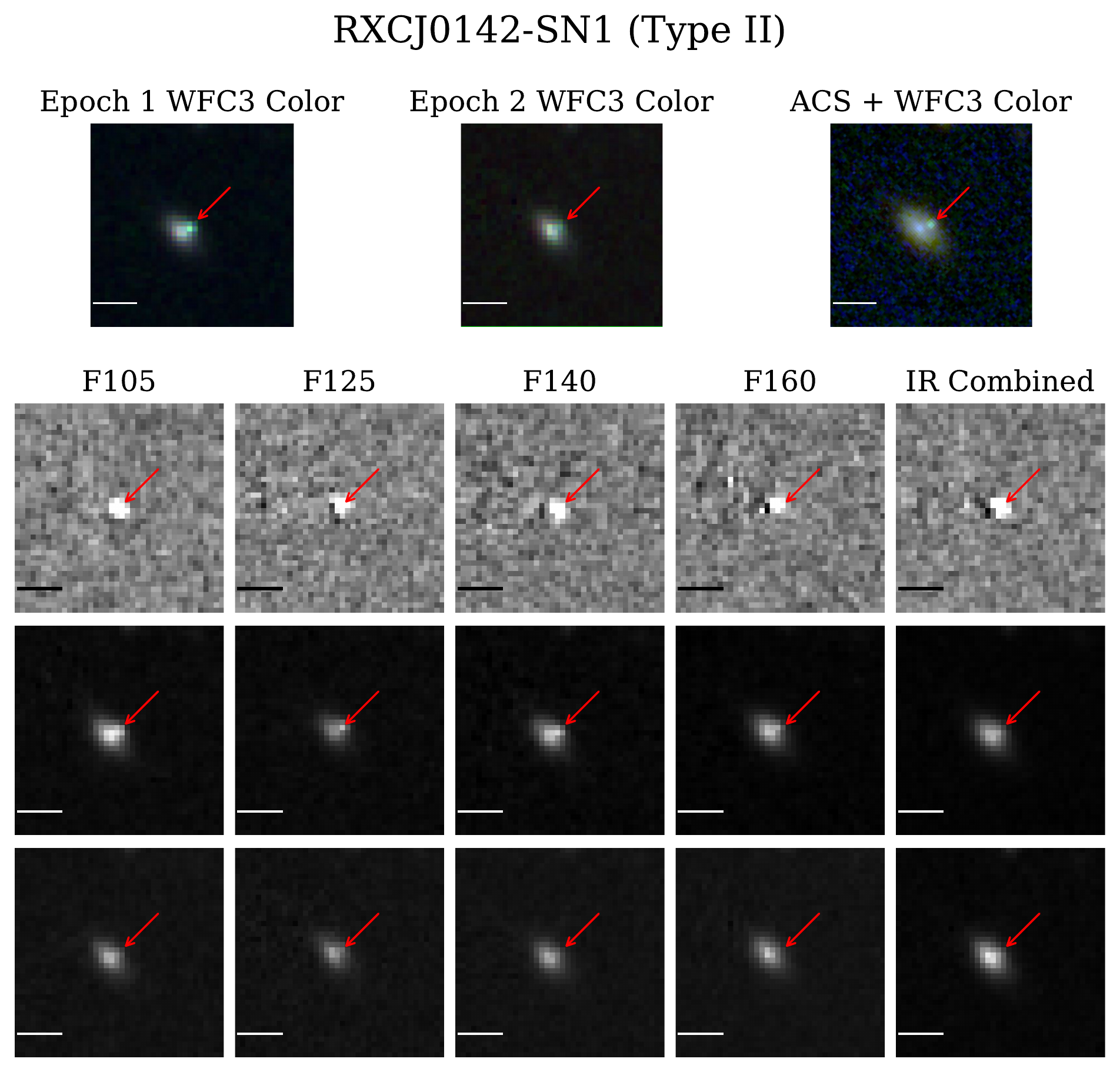}
  \vspace{0.5cm}
    \includegraphics[height=0.40\textwidth,width=0.49\textwidth, keepaspectratio=true]{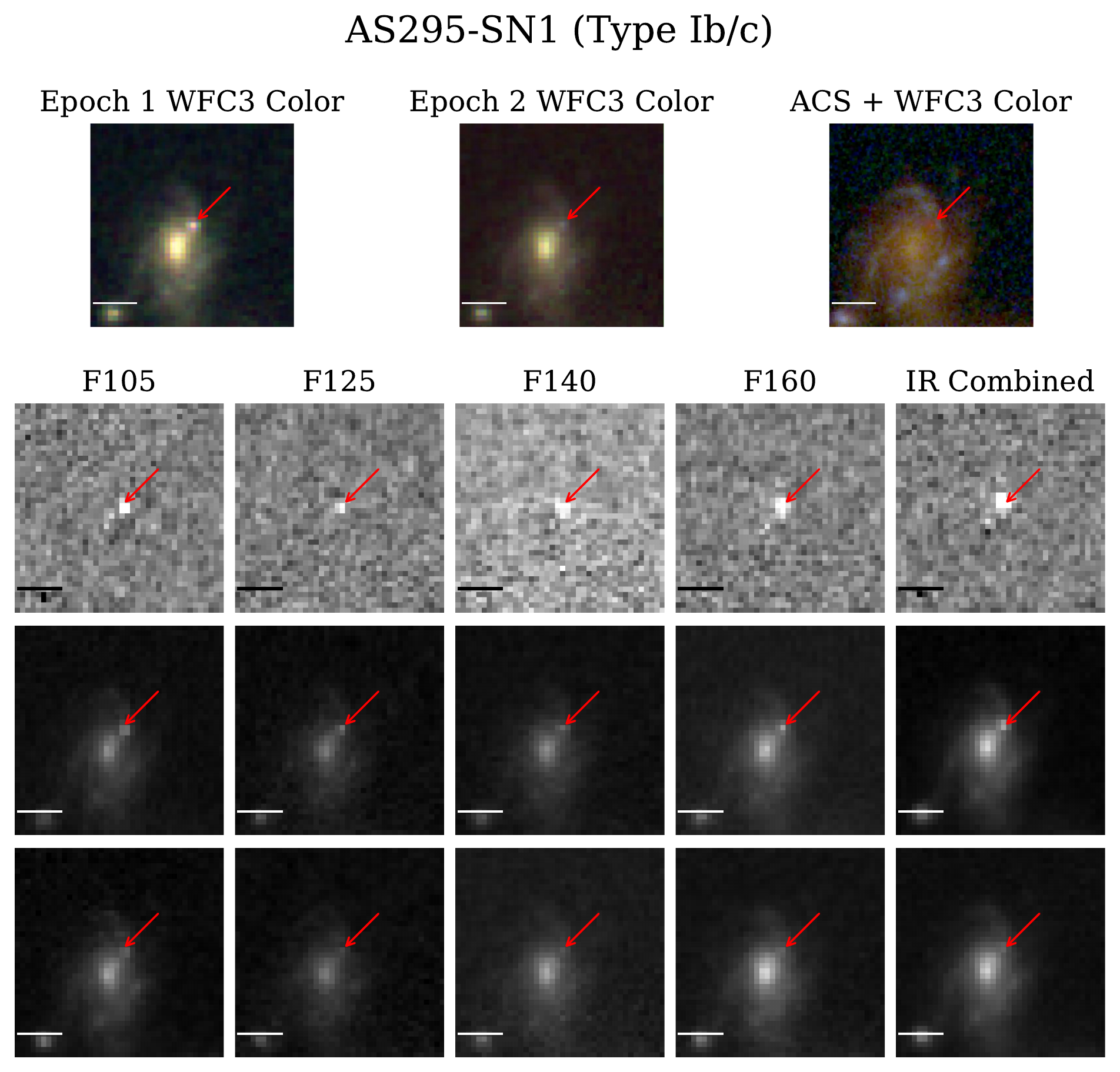}
    \includegraphics[height=0.40\textwidth,width=0.49\textwidth, keepaspectratio=true]{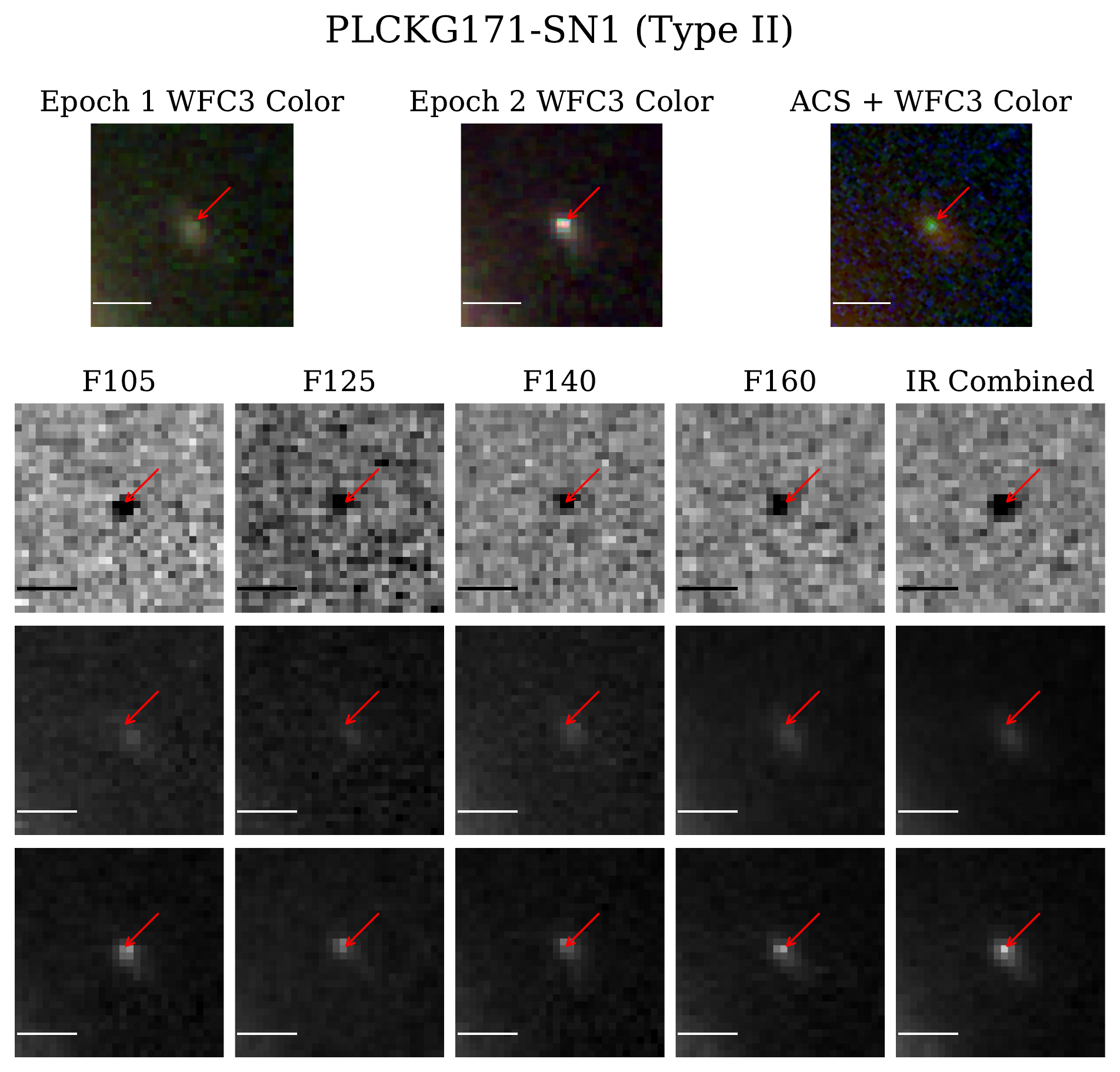}
  \vspace{0.5cm}
    \includegraphics[height=0.40\textwidth,width=0.49\textwidth, keepaspectratio=true]{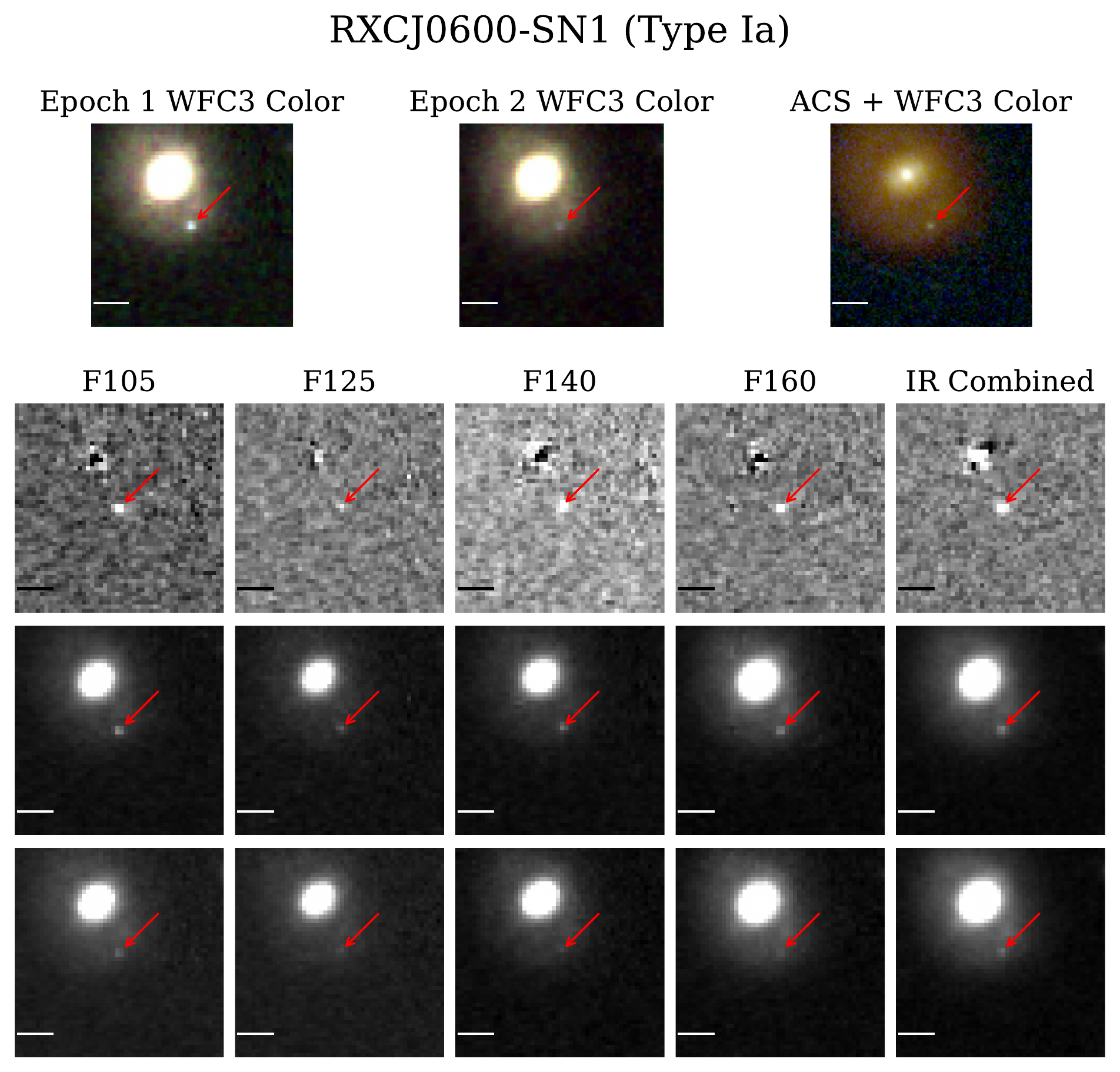}
    \includegraphics[height=0.40\textwidth,width=0.49\textwidth, keepaspectratio=true]{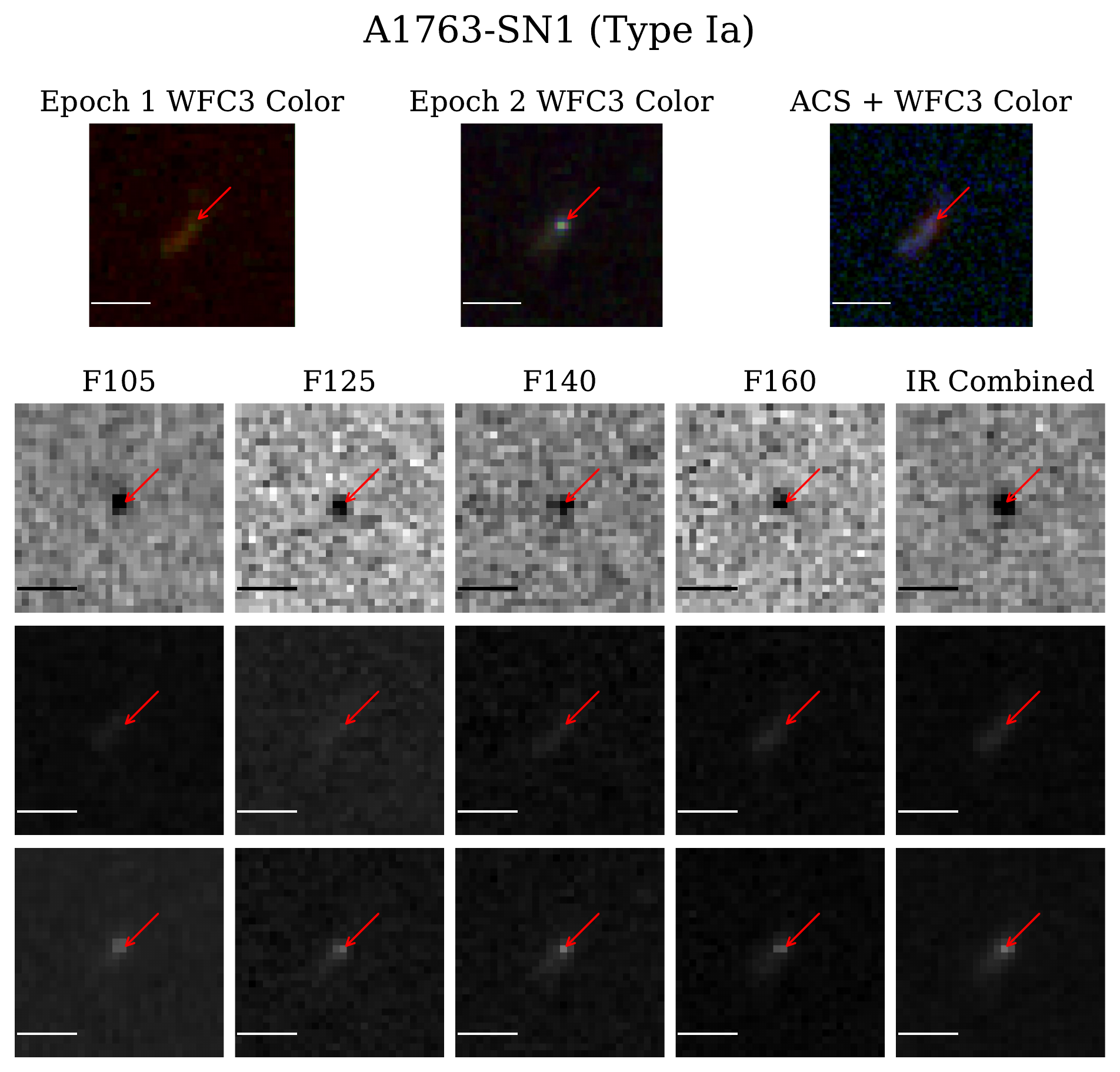}
    \includegraphics[height=0.40\textwidth,width=0.49\textwidth, keepaspectratio=true]{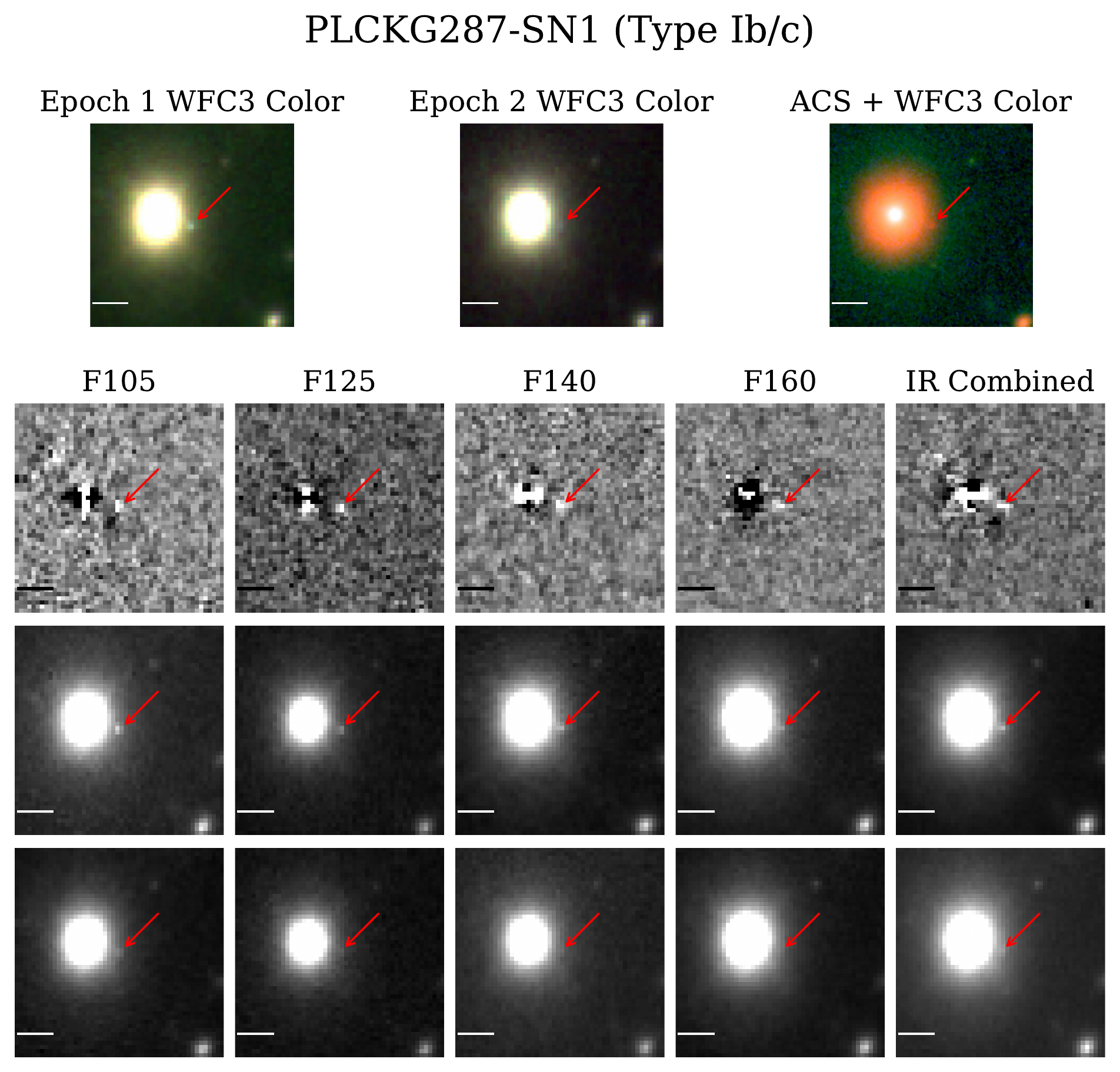}
 \end{center}
\caption{The six SN detections. For each one, at the top row we present three colour images, one for each epoch from the WFC3 filters, and one including all ACS+WFC3 filters; Second row displays the difference image for each filter; Next two rows present the images from the two epochs. A scale of $1\arcsec$ is marked upon the image. The red arrow in each stamp points to the exact candidate location. In parentheses we designate for each SN candidate the best fitted type. The orientation of the figures is arbitrary.}\vspace{0.1cm}
\label{fig:curve_all}
\end{figure*}
\clearpage
\section*{Data Availability}

 The data used in this work are publicly available on the MAST archive  and the RELICS website.


\bibliographystyle{mnras}
\bibliography{example} 








\bsp	
\label{lastpage}
\end{document}